# Numerical Assessment of Advective and Diffusive Dynamics of Interacting and Isolated Prototypical Convectively Initiated Circulations

Matthew R. Igel[a]*, Joseph A. Biello[b], and Adele L. Igel[a]

[a]Department of Land, Air and Water Resources, University of California Davis, Davis, USA; [b]Department of Mathematics, University of California Davis, Davis, USA

*corresponding author: migel@ucdavis.edu

# Numerical Assessment of Advective and Diffusive Dynamics of Interacting and Isolated Prototypical Convectively Initiated Circulations

The bulk circulation associated with convective clouds includes not only a region of updraft and cloudy air but also a region of compensating descent and cloud-free air and horizontal motions coupling these regions. The Kinematic Representation of Non-rotating Updraft Tori (KRoNUT) model is a simple representation of this entire flow. First, the skill of the KRoNUT in representing flows from a high resolution full-physics simulation of marine tropical convection is compared to various plume representations of convection. Then the KRoNUT is used to construct bulk descriptions of the dry dynamics of isolated and interacting convective circulations under the influence of advection and diffusion (only). Cross sections of advective and diffusive tendencies show that while vertical advection of the vertical wind is the most important advective tendency in clouds, the horizontal component of the convective circulation and advection thereof plays a crucial role in the evolution of circulations in the absence of buoyancy. Strong curvature of the flow near the surface and near the updraft core results in locally strong diffusive tendencies that depend on scale. Cross sections of tendencies from the KRoNUT compare favourably to results from the simulation. Interacting circulations are shown to exhibit a wide range of dynamics with some cases of interactions leading to unique stability of geometric properties of otherwise evolving flows and some leading to geometric clustering of circulation centers.

**Introduction**

Completely understanding the evolution and behaviour of convective clouds has proven to be an elusive goal (Arakawa 2004; Bony et al. 2015). Many obstacles stand firmly in the way; several outstanding questions must be answered but seem impossible to be. For example, how do myriad environmental properties couple to latent heat release through complex microphysics? Or, how do simple parcel-level processes extend in time and space to result in the observed momentum distribution in clouds? These questions may even lack general answers.

Thus, understanding convective clouds is and likely always will be a challenge. Our suggestion for how to make progress is to introduce a not-too-complex but grossly representative model for convection. Our model sits between a parcel, which is a local in space and time model, and a plume, which integrates over time and space to form a steady state. Our model, the Kinematic Representation of Non-rotating Updraft Tori (KRoNUT) (Igel and Biello 2020) (see section 2 below), takes the shape of a torus and includes both rapidly rising updraft air and slowly descending subsident air. But unlike most fundamental model implementations of plume theory (Arakawa and Schubert 1974) existing in a box of single-valued (at any height) subsidence, the KRoNUT has horizontal structure in both its updrafts and subsidence and can be made to evolve. The KRoNUT is not a facsimile of convection. Rather, it is a time-space approximation of the flow from which we might hope to begin to tackle stubborn questions regarding convectively initiated circulations.

Among those intractable problems is the dry, neutrally buoyant, incompressible dynamics of convective clouds. In a companion paper (Falcone et al. 2025) (FIB25 hereafter), we asked: how does a KRoNUT flow evolve in time due to advection and diffusion (i.e. dry components of its dynamics) in isolation and when affected by another nearby KRoNUT flow? FIB25 took an analytic approach. The benefit of their method was elegance, tractability, and clear outcomes. But the complexity of their method left little room for numerical testing and did not provide key details about how bulk changes to circulations were driven by local tendencies below the cloud scale. So, here we ask *what brings about* the evolution of KRoNUT flows due to dry dynamics in isolated and coupled circulations? One of the unique results of FIB25's approach was the ability to describe a unified goal of the dry dynamics of circulations. They showed that all circulations evolve toward a state of zero circulation tendency. Some clouds achieve this goal through weakening in time while others do so by adjusting toward a specific geometry.

While obeying that global goal, FIB25 made several suggestions about clouds. One is that clouds will always get taller in time due to advection and diffusion and that most cloud morphologies will grow vertically faster than they grow horizontally. Another is that for a given diffusion, short, wide clouds will tend to behave very differently than other clouds (in a way we will describe below). Finally, FIB25 suggests that the range of behaviours available to KRoNUTs representative of shallow circulations is much wider than deep convective clouds.

Of course, focusing on advection and diffusion ignores important dynamic contributions from dynamic pressure perturbations (List and Lozowski 1970; Davies-Jones 2003; Peters 2016) and, more obviously, latent heating. But, integrating buoyancy forcing necessitates coupling the KRoNUT to thermodynamics. Doing so is conceptually challenging but is an area of active work by the authors. Therefore, the description below represents a useful starting point and not a full description of the dynamics of clouds. Most of our discussion is done in a scale agnostic way (except infrequently in a narrative manner) so as not to imply that the dynamics we discuss are more than a portion of the true dynamics. Still, it is worth drawing attention to the fact that being able to assume this scale agnostic discussion illustrates the utility of the KRoNUT framework rather than a drawback. Finally, we want to point out that our discussion below is not exhaustive of all existing conceptual models of convection which range in time and space scales from parcels, to Hill's vortices, to plumes. We focus on comparing the KRoNUT to the plume.

This paper is organized in a way that allows us to introduce some basic tools, test them, and examine fine-scale details as follows. First, we discuss models for convective motion and how we identify parameters of these models in general. Then, we discuss identifying convectively initiated motion in a high-resolution simulation and contrast the relative skill of circulation models. Next, we examine the behavior of advection and diffusion in convective circulations in a way that has not been possible before. Finally, we return to the simulations to further test our tools.

**The KRoNUT and Static Plumes**

*The KRoNUT Model*

A thorough reintroduction of the KRoNUT is provided in FIB25 and will not be provided here. The model consists of a divergence-free velocity field with azimuthal symmetry. The horizontal (i.e. radial, *u*) and vertical (*w*) wind components are

$$u(r,z) = \frac{-w^* r}{2H}(1 - \frac{z}{H})e^{1-z/H-r^2/L^2} \qquad (1)$$

$$w(r,z) = \frac{w^* z}{H}(1 - \frac{r^2}{L^2})e^{1-z/H-r^2/L^2} \qquad (2)$$

where *L*, *H*, and $w^*$ are parameters of the KRoNUT most easily understood by examining Fig. 1. *L* is the radius at which *w(r,z)* becomes negative at all heights. *H* is the height at which the vertical wind speed peaks. $w^*$ is the magnitude of the maximum vertical wind (at *H*).

By construction, the KRoNUT does not differentiate between updrafts and downdrafts or subsidence. Rather, both are part of a single circulation. Figure 1a illustrates the basic properties of the KRoNUT: upward flow inside of ring of radius *L* and downward beyond; inward flow below *H* and outward above. Although the vertical velocity decays rapidly with height, Fig. 1 highlights that the flow is unbounded in *z*, which is a notable limitation of the current version of the KRoNUT. Adding a semi-rigid top to the flow in a way motivated by full-physics simulations is ongoing work. The green vectors are included in Fig. 1a to illustrate that given only vector wind, it would be easy to mischaracterize the KRoNUT flow as being comprised of an updraft only. We want to draw a parallel between that potential mischaracterization and the one we feel we often make when observing real clouds which may have strong updrafts and high liquid water content, but which are not isolated from the fluid around them despite the appealing simplicity of that mental picture. Subsiding shells and broader subsidence are important elements of the real cloudy circulation (Heus and Jonker 2008).

*KRoNUT Parameter Finding*

Because any KRoNUT has three parameters, $L$, $H$, and $s^*$, identifying a KRoNUT in any arbitrary flow requires three metrics plus a method to identify a location in space. For 3-dimensional data, we first identify the horizontal center of a potential KRoNUT circulation by finding horizontal locations of local maxima in vertically summed vertical velocity. Then, we azimuthally average the vertical wind and the projection of the full horizontal wind on rays from the identified center. These averages are then projected onto an isotropic radial-vertical grid. All subsequent calculations are performed with these cylindrical coordinates. Next, we compute three spatial moments of the vertical kinetic energy as

$$[M_1, M_r, M_z] = \frac{1}{2}\int [1, r, z] w^2(r, z) dV \qquad (3)$$

to match the number of independent model parameters where $dV$ is the cylindrical volume element. These three moments, the $1^{st}$, $R^{th}$, and $Z^{th}$, are compared to a lookup table of pre-computed moments for various KRoNUTs which were calculated on the same grid topology as the target flow. We then find the combination of parameters which minimizes the sum of the absolute value of the logarithm of the ratio of the flow's moments to the pre-computed moments. Using the ratio of moments helps to normalize for the inherent difference in the magnitude of the various spatial moments, and using the absolute value of the logarithm allows us to search for a minimum (rather than a particular value of the sum).

To test the fitting method, we computed KRoNUT velocity fields for geometries selected at random from among those used to construct the lookup tables. To the velocity field of each KRoNUT we applied a spatially correlated noise field by computing a moving average over 2500m to a normally distributed noise distribution with a mean of one. An example vertical velocity field with this noise is shown in Fig. 2a. While the underlying structure of a smooth KRoNUT velocity field is apparent, the actual small-scale structure of the velocity is quite noisy, and it is not obvious that a moment calculation will necessarily represent the underlying smooth structure well. Nonetheless, moments of this noisy field were calculated and used to fit parameters through the normalization described in the paragraph above. This procedure was done 10,000 times. The resulting distributions of percent errors in fitting for $L$, $H$, and $w^*$ are shown in Fig. 2b-d. Errors are normally distributed. The median error for $L$ and $H$ is 0 m, while the median error in $*^s$ is 0.10 m s$^{-1}$ (about a 1% error). The small positive error in fitting $*^s$ is likely due to the fact that the application of random noise to $w$ slightly raises the total vertical kinetic energy, the quantity we use for fitting. The standard deviation of errors in $H$ is 50% greater than that of $L$. Qualitatively, it appears the method has the hardest time fitting $w^*$ accurately. The example in Fig. 1a is a reasonable illustration of this. The application of noise raises the maximum velocity by over 4 m s$^{-1}$ and the fitting diagnoses a $w^*$ that is 0.22 m s$^{-1}$ (+2% error) too high, but the diagnosed *L* and *H* are off by only -100 m (-1% error) and 100 m (1% error), respectively. The latter two errors are smaller than the grid spacing of 250 m.

***Plumes***

Several forms of plume will be used as a foil for the KRoNUT in section 3. To contrast a plume with a KRoNUT fairly, we seek simple representations of a plume with the same number of parameters as a KRoNUT. These are a width, L, a height, H, and an intensity, w. We test four reasonable plume types formed from the product of two binary geometric options: i) the plume may be of constant width, L, from its bottom to top at height H (a "box"), or it may be of zero width at its bottom and a final width, L, at its top at H (a "cone"); ii) the plume may have a radial "top hat" vertical velocity profile or a Gaussian one with a vertical velocity profile clipped to zero at L which is two standard deviations from the core (a "bell"). Subsidence outside the plume is radially constant to a radius of 20 km but vertically variable to exactly offset the total upward motion. These plumes are designed to span possible characteristics suggested by the wide range of early literature on the subject (Garvey and Fowler 2023).

**Plume Parameter Finding**

Plume diagnosis follows the KRoNUT diagnosis procedure as closely as possible. We diagnose plumes by minimizing a total difference (as above with KRoNUTs) of the $1^{st}$, $R^{th}$, and $Z^{th}$ moments of the vertical kinetic energy of identified circulations to pre-computed circulations like we do with the KRoNUT. While minimizing azimuthal vorticity moments provide an excellent means of identifying KRoNUTs (not shown here but used in FIB25), there is no reasonable horizontal velocity in plumes with locally discontinuous vertical velocity and so no reasonable vorticity. To identify KRoNUTs and plumes on equal terms, we use moments of the vertical kinetic energy for both although this significantly undervalues the KRoNUT which has significant information tied up in its horizontal flow. And while used as a good enough numerical identification technique here, kinetic energy would fail to produce the kind of closed form approximations seen in FIB25 (and citations therein using vorticity) without additional physical assumptions.

**KRoNUTs and Plumes in DYNAMO**

To this point in the KRoNUT's brief history, its similarity to data has not been shown. While the KRoNUT will always necessarily be a gross simplification of any complex real-world convectively initiated circulation, it would be useful to have some confidence that it reflects important aspects of observed flows reasonably. Full 3-dimensional velocity fields of the cloudy and clear air components of convective circulations across scales are not observed in large numbers in any data set of which the authors are aware. So, we use numerical simulations of the tropical Indian ocean during the Dynamics of the Madden-Julian Oscillation (DYNAMO) campaign.

*Simulation Description*

To simulate a large field of deep convective elements, we used the Regional Atmospheric Modeling System (RAMS) (Saleeby and van den Heever 2013) to simulate one day during an MJO event in the Indian Ocean during the DYNAMO field campaign (Yoneyama et al. 2013; Zhang et al. 2013). We simulated a 16-hr period beginning 22 November 2011 at 12Z. Initial conditions came from the ERA5 reanalysis (Hersbach et al. 2020). This data product was also used to nudge the simulations within 30km of the lateral boundaries. The domain was 1800 km x 3000 km x 25 km and centered at 75ºE at the equator. Grid spacing was 3 km in the horizontal and variable in the vertical with a minimum spacing of 25 m just above the surface stretching to a maximum of 250 m. A four second time step was used. No convective parameterization was used, the RAMS two-moment microphysics scheme was employed, sub-grid diffusion was parameterized following Smagorinsky (1963), the RTE-RRTMGP radiation scheme was used (Pincus et al. 2019), and the ocean surface was assumed to have a temperature of 302.7 K. This simulation was used to initialize and nudge a higher resolution simulation on a subdomain that contained many deep convective cores. This second simulation started at 00Z on 23 November and ran for 4 hours. It was centered at -2.7123°S and 73.7710°E. It used a 300 km x 400 km x 25 km domain with 250 m horizontal grid spacing and a two second time step. All parameterizations were the same as for the parent simulation.

Because the KRoNUT represents a sort of broad, mean circulation, 5-minute average DYNAMO output is interpolated and saved on an isotopic 250 m grid across the whole simulation domain for the final 3 hours of the simulation period. Figure 3 shows the result of our decision to analyze data on the averaged grid in contrast to doing so on the native grid of the model at a single timestep. The averaged flow is more laminar and more upright than the output from the single timestep. We believe this is an important point. The time and space averaged data represent well the macroscale features of the convection occurring but sufficiently smooths the velocity field in a way that removes transient features that may overcomplicate utilizing the data. This is important for the KRoNUT because it means that it is relevant to describing mean flows on timescales on the order of several minutes, rather than instantaneous or mean whole-lifetime flows, say. As a note, the location shown in Fig. 3 was chosen objectively; it is the first identified candidate location for an updraft core found using the method outlined above.

*Extraction of Simulated Circulations*

We extract circulations from the simulations using the radial compositing and moment method described above. First, the candidate circulation geometry and intensity are found. Then, the wind field implied by the parameters identified is subtracted from the full simulation wind field, and the process is repeated 50 times per output time (this is a somewhat arbitrary choice that worked well in practice). Doing this with a reasonable geometric representation of a circulation would steadily remove updraft kinetic energy from the simulation wind field.

Figure 4a shows the incremental extraction of vertical kinetic energy from the simulation output fields after 50 circulations are removed. All morphologies remove vertical kinetic energy except for the hat cone which for some output velocity fields results in an increase due to poor diagnosis of the flow. The bell box, which is the most KRoNUT-like of the plumes, removes the most kinetic energy in the mean while the KRoNUT removes the second most in the mean and most for some output periods. We cannot prove the KRoNUT is better than any plume, but we certainly fail to reject the null hypothesis that it is worse and therefore not worth exploring by this test.

Figure 4 also includes a scatter plot of KRoNUT fit values. These are included to provide a sense of scale. We see a range of mesoscale $H$ and $L$ with little correlation between the geometric parameters and a loose inverse relationship between $w^*$ and both $L$ and $H$. Approximately 9% of all $L$ values are equal to the maximum included in the table of 18 km.

**Advection and Diffusion of a Single KRoNUT**

Next, we move on to exploring the nature of advection and diffusion in circulations.

*Advection*

*Cross Sections*

As we've previously stated, one of the benefits of the KRoNUT is that the circulation is fully three-dimensional. This allows us not only to derive advective tendencies from the $u$ and $w$ flows with themselves, but also with each other. The four advective terms are

$$-u\frac{\partial u}{\partial r} = \frac{-u^2}{r}(1 - 2\frac{r^2}{L^2}) \tag{4}$$

$$-w\frac{\partial u}{\partial z} = \frac{uw}{H}\frac{(2 - \frac{z}{H})}{(1 - \frac{z}{H})} \tag{5}$$

$$-u\frac{\partial w}{\partial r} = \frac{2uw}{L}\frac{(2 - \frac{r^2}{L^2})}{(1 - \frac{r^2}{L^2})} \tag{6}$$

$$-w\frac{\partial w}{\partial z} = \frac{-w^2}{z}(1 - \frac{z}{H}) \tag{7}$$

Before discussing these, we'll point out that the form of Eqs. (1) and (2) allow the advection terms to be expressed as products of the velocities themselves multiplied by various geometric factor terms. This is not by design (precisely) but is a convenient consequence. And as a simple check, we notice that the RHS of each advection term is a product of velocities divided by a length which yields units of acceleration.

Equations (4) and (7) show that the sign of "self-advection" of $u$ only depends on $r$ and that the sign of $w$ only depends on $z$. So, $u$ tendency from self-advection is inward near the core ($r < \frac{L}{\sqrt{2}}$) and outward toward cloud edge ($r > \frac{L}{\sqrt{2}}$) and in the subsiding air at all levels (Fig. 5a). Similarly, $w$ tendency is downward below the level of maximum updraft and upward above at all radii. Together, self-advection acts to narrow and raise the updraft core.

Equations (5) and (6) represent "cross-advection" from a KRoNUT acting on itself. Equation (5) implies that *w* tends to force inward motion within the updraft at most heights and outward motion in the subsident region. This tendency is diffluent up to *z=2H* and is at least as large as self-advection at heights near $z \approx \frac{H}{2}$. The cross-advection of *w* by *u* is weak relative to self-advection and acts to amplify the tendency from self-advection out to $r=\sqrt{\frac{3}{2}}L$. The magnitude of Eq. (6) peaks at $r = \frac{1}{2}(\sqrt{5}L - L)$ o, $r \approx \frac{2}{3}L$ where it is locally greater than the self-advection term.

The total *u* and *w* tendencies from advection are shown in the bottom row of Fig. 5. Tendencies are largest within the cloud ($r < L$) and especially below the level of maximum updraft ($z < H$) where curvature of the flow is high. The *w* tendency mostly echoes its self-advection term except as *r* approaches *L* where cross-advection dominates. Intriguingly, the tendency of *w* at low levels is negative within the ascent region. Low-level forcing for descent is usually attributed to cooling and loading by hydrometeors in clouds, but here, we present a purely advective mechanism (similar to (Heus and Jonker 2008)). The reflection of this is that there is upward advection above *H* in the decent region beyond *L*. Total advective tendency of *u* is weaker than *w*. There is a net enhancement of inflow at low *r* within most of the ascent region but enhanced outflow at nearly all heights for $r \approx L$ and for all *r* at very high heights.

*Numerical Assessment.*

The KRoNUT represents a kind of "bulk" model of a convectively initiated circulation – a whole 3-dimensional convective flow is encoded in three simple model parameters. FIB25 presented tendencies of these bulk parameters due to advection and diffusion for both a single KRoNUT and a pair of interacting KRoNUT circulations. The FIB25 method is analytic but still a bulk method. Here, we would like to develop tendencies from a smaller scale perspective.

To do so, we compute tendencies of velocity due to physics on a simple radial-vertical grid assuming azimuthal symmetry. Grid spacing is a uniform 250m. The range of $r$ and $z$ is [0, 20 km]. Both match the spacing of pre-computed moment tables used above. On this grid, we compute tendencies from advection and diffusion computed from the velocity field with simple centered differences for derivatives. We employ a constant density and pressure fluid for simplicity and for consistency with FIB25 assumptions. We initialize the grid with flows for a wide range of $L$ and $H$. This range is 1 km<$(L,H)$<10 km. The tendency of parameters is the difference between the diagnosed parameter after a physics timestep divided by the timestep length. We average the tendencies implied by employing timesteps of between 300 s and 600 s spaced by 100 s. We also average over $w^*$ from 3 m s$^{-1}$ to 8 m s$^{-1}$ to add robustness.

The left column of Fig. 6 shows the tendencies from advection of bulk parameters using this method. $\dot{H}$ is always positive. $\dot{L}$ is largely negative for most $H$. So, even without buoyancy, most circulations grow upward but contract horizontally. Figure 5 and Eqs. (4)-(7) can shed light on why. It would seem that $\dot{H}$ is always positive because total advection of $w$ is always positive above the level of maximum vertical wind. $\dot{L}$ is more difficult to intuit. The total tendency of $w$ would tend to act to widen the updraft above $H$. The self-advection of $u$ above $H$ would tend to force wider KRoNUT-like circulations.

For high $L/H$, $\dot{L}$ is intensely negative. After examining maps of advective tendencies in this KRoNUT phase space, the most consistent difference with tendencies elsewhere in the $L$-$H$ phase space is not the structure of the tendencies but the magnitude. In this region, the curvature of the flow is extremely high. Magnitudes of advective tendencies for the same $w^*$ are much higher for high $L/H$.

In FIB25, $\dot{w}^*$ was shown to be positive when $L/H>1.724$. Our numerical method here depends mostly on $H$ (Fig. 6e). Shallow circulations uniformly speed up while deep circulations uniformly slow down very slightly. One should note that $\dot{H}$ and $\dot{w}^*$ need not be of the same sign. That is, shallow KRoNUT strengthen and grow vertically while deep KRoNUT weaken yet still grow vertically (because $w^*$ is still positive).

## *Diffusion*

### *Cross Sections*

Next, we can present the shape of diffusion components for the velocity tendency.

$$\frac{1}{r}\frac{\partial}{\partial r}\left(r\frac{\partial u}{\partial r}\right) - \frac{u}{r^2} = \frac{-4u}{L^2}\left(2 - \frac{r^2}{L^2}\right) \tag{8}$$

$$\frac{1}{r}\frac{\partial}{\partial r}\left(r\frac{\partial w}{\partial r}\right) = \frac{-4w}{L^2}\frac{\left(2 - 4\frac{r^2}{L^2} + \frac{r^4}{L^4}\right)}{\left(1 - \frac{r^2}{L^2}\right)} \tag{9}$$

$$\frac{\partial^2 u}{\partial z^2} = \frac{u}{H^2}\frac{\left(3 - \frac{z}{H}\right)}{\left(1 - \frac{z}{H}\right)} \tag{10}$$

$$\frac{\partial^2 w}{\partial z^2} = \frac{-w}{zH}\left(2 - \frac{z}{H}\right) \tag{11}$$

The RHS of Eqs. (8)-(11) have been rearranged to be functions of velocities (like the advection components) so that it is easy to see that they have units of velocity divided by two lengths, or when multiplied by a viscosity, $v$, acceleration. For numerical calculations, we use a $v=100$ m$^2$ s$^{-1}$; however, our choice is only consequential to the magnitude of results shown not their shape. Interestingly, the denominator of the of Eq. (11) is a product of both $H$ and the physical dimension, $z$. This means that, unlike in the case of advection, diffusion is not self-similar for all KRoNUT.

Figure 7 shows cross sections of diffusive terms. Both horizontal (Fig. 7a) and vertical (Fig. 7c) diffusion of the radial wind are strongest near the surface about halfway between the cloud core and its edge. Interestingly, horizontal and vertical diffusion are similarly shaped but of opposite sign (apparent comparing Eqs. (8) and (10)). The total diffusive tendency of the horizontal wind is outward for $r<L$ and $z<\sim\frac{2H}{3}$ and inward elsewhere (or essentially zero). This would force a confluent tendency near the surface near the updraft edge at $r=L$.

Within much of the cloudy updraft of the KRoNUT, $w$ is weakened by both vertical and horizontal diffusion. However, there is a region just within the updraft to just within twice the updraft width ($\sqrt{(2-\sqrt{2})}L < r < \sqrt{(\sqrt{2}+2)}L$; i.e. roots of Eq. (9)) at all heights and out to high $r$ near the surface where the tendency of $w$ is positive. So, both the regions around the updraft-subsidence interface as well as the boundary layer far outside the circulation center are diffused upward.

Also in Fig. 7 are the total diffusive tendencies of $u$ and $w$ for a short and wide circulation. Equations (8) and (10) suggest that when $L/H$ is large the nature of the cancellation of vertical and horizontal diffusion of the horizontal wind may change. The conclusions drawn from maps about the nature of diffusion here are insensitive to the relationship of $L$ and $H$ (mostly not shown) except in this case when $L/H$ is large. In this case, vertical diffusion becomes very important such that horizontal winds are diffused inward at low levels at all relevant $r$ and upward diffusion of vertical winds extends inward toward very low $r$ at moderate heights. Thus, low-level inflow and upper-level updrafts are reinforced for these circulations by diffusion.

*Numerical Assessment*

Bulk diffusive tendencies are shown in Fig. 6 (right column). The least noisy result is that $\dot{w}_s$ is uniformly negative. That is, the intensity of the circulation decreases because of diffusion. It does so most strongly for small $L$. This seems to be a result of a widening of the circulation (i.e. positive $\dot{L}$) at small $L$ (Fig. 6d). The behaviour of $\dot{H}$ is disorganized but largely positive. So, in the absence of any kind of thermodynamic forcing, diffusion tends to weaken convective circulations while making them taller. Combined with the results from Fig. 7, the diffusion of the vertical wind perhaps plays a bigger role in determining the bulk tendencies than the diffusion of the horizontal wind. The diffusion of the vertical wind would clearly tend to weaken the upward motion of the circulation while potentially also acting to grow it horizontally thanks to the region of upward forcing at and beyond updraft edge throughout the depth of the troposphere. These results reflect those in FIB25 who likewise suggest geometric growth and enervation from diffusion except for very high $L/H$ circulations.

*Cross Sections from RAMS*

To conclude our discussion of the dynamics of a single KRoNUT, we return to the RAMS simulations of the DYNAMO case. We use the identified KRoNUT locations in the simulations to construct composites of kinematic flows and physical tendencies. In addition to outputting the 5-minute averaged kinematic fields on an interpolated grid, we also recorded the physical tendencies which we use to compare to KRoNUT derived tendencies. Note that these simulations include full diabatic physics and parameterized turbulence.

The first row of Fig. 8 shows composite (azimuthally interpolated) $u(r,z)$ and $w(r,z)$ on a grid which interpolates results to the identified $L$ and $H$ on a case-by-case basis before averaging all the cases. The vertical wind composite shows a clear updraft which peaks just above $H$ and which switches sign slightly beyond $L$ (Fig. 8a). Details, like a downdraft below the updraft and some waviness near $z=1.9H$, are apparent and not part of the current form of the KRoNUT. The downdraft also affects the inflow (Fig. 8b) as it is forced to rise above the surface as it approaches the convective core. The horizontal wind consequently switches sign slightly higher than $z=H$ at low $r$ and slightly lower at high $r$. The depth of the upper layer outflow is also more confined than it is in the KRoNUT. This stems from the lack of a semi-rigid tropopause in the current version of the model.

The remaining panels of Fig. 8 show composite physical tendencies of the radial and vertical wind from the model's advection and diffusion. The chief conceptual limitation to showing these maps is that they account for the total advection and diffusion not just the contribution from the extracted KRoNUT circulation (but due to the total extracted flow which may include other small-scale circulations that do not average out in the radial compositing). Because of the elevation of the inflow off the surface, the total advection of the radial wind exhibits a markedly different structure than the KRoNUT below $z=H$. Advection is outward everywhere rather than convergent near the surface below $z=H$ but is negative near the convective core above $z=H$ (Fig. 8c). Diffusion of $u$ is an order of magnitude weaker than advection (Fig. 8d). Like the KRoNUT, diffusion within $r=L$ is outward-below-inward and opposes the background flow while beyond $r=L$, it reinforces the background. Advection of the vertical wind is upward-over-downward near the core and exhibits the change in sign suggested by the horizontal advection in the KRoNUT (Fig. 8e). Diffusion is downward in the convective core and noisy outside although mostly positive from $z=0.4L$ to $z=L$ (Fig. 8f). Finally, we'll note vertical tendencies are an order of magnitude greater than horizontal tendencies, like in the KRoNUT. Overall, these dynamic tendencies exhibit important similarities to those implied by the KRoNUT but also consequential differences.

**Advection and Diffusion of Two KRoNUT Circulations**

Real-world cloud-cloud interactions are known to be complex and highly varied, and critical to the mesoscale evolution of the cloud field. Clouds of various types and morphologies interact directly through their primary circulations (Chen et al. 2023) and indirectly through their secondary circulations (Haerter 2019; Rotunno et al. 1988; Stevens et al. 2005), modification of the convective thermodynamic environment (Wing and Emanuel 2014; Wing et al. 2017), and through gravity wave excitation (Mapes 1993; Lane and Zhang 2011). The KRoNUT provides a means of better understanding direct kinematic interactions which are difficult to observe in nature and difficult to assess quantitatively in a full physics simulation.

Here, we embed one KRoNUT within the circulation of a second so that the first KRoNUT circulation evolves due to the flow of the second. The evolving KRoNUT and its parameters will be given a subscript 'e' (e.g. $KRoNUT_e$, $L_e$, etc.). The static KRoNUT and its parameters will be given a subscript 's'. The total tendency on $KRONUT_e$ due to advection and diffusion can be approximated by summing its tendencies due to interaction with $KRoNUT_s$ with the tendencies presented in section 4. FIB25 provides details on how the tendencies sum in the slightly more general case of two evolving circulations; they use subscripts 'n' and 'f' to track two evolving KRoNUT's.

*Dynamics Maps*

To perform these calculations, we use a Cartesian grid. Azimuthally symmetric

KRoNUT flows are interpolated onto a uniform grid with spacing of 250 m. KRoNUT$_e$ is centered on the 20 km cube domain whereas KRoNUT$_s$ is offset from the center by a distance $\Delta s$. KRoNUT$_s$ is then advects or diffuses KRoNUT$_e$. After the evolution, the maximum summed vertical velocity is used to find the location of the KRoNUT$_e$ center. The flow is then azimuthally averaged about the new center. The new KRoNUT$_e$ parameters are then found as above. An example of the two-KRoNUT total vertically summed vertical wind and the difference after an advective and diffusive step are shown in Fig. 9. One can see the two KRoNUT centers in panel a). KRoNUT$_e$ has $w^*_e$=5 m s$^{-1}$, $L_e$=3 km, and $H_e$=3 km. KRoNUT$_s$ has $w^*_s$=10 m s$^{-1}$, $L_s$=5 km, and $H_s$=8 km. We show separations, $\Delta s$, of 3 km and 8 km so we can discuss differences for initially independent circulations within and outside the updraft of another even if the former yields a somewhat awkward picture of a cloud partially within another.

The lower two rows of panels show the vertical mean tendency from advection and diffusion. While showing a vertical mean can only provide a gross view, a few things are apparent. Both advection and diffusions act to enervate and invigorate the local vertical wind. These physics act asymmetrically on the recipient KRoNUT$_e$. Advection tends to excite updraft between the two KRoNUTs. When $\Delta s$ is small, the updraft tendency is circular, and a crescent shaped downdraft tendency exists from $x \approx$ 0 km to $x \approx$ 5 km. When $\Delta s$ is large, the updraft tendency region is elongated in $y$ about halfway between the two updraft centers. Downdraft tendency exists on either side of this region including in the core updraft region of KRoNUT$_e$ at ($x$=0, $y$=0). Regardless of $\Delta s$, advection tends to accelerate air upward in the far field. The diffusive tendency has the same shape and magnitude regardless of $\Delta s$ because it only depends on KRoNUT$_s$. Diffusion creates downdraft tendency at the KRoNUT$_s$ core, outward to just beyond $L_s$. Then it is ringed by updraft tendency and then by downdraft in the far field.

The nature of these interactions is sensitive to KRoNUT parameters, and we cannot hope to show all possible kinds of interactions. Rather, Fig. 9 is meant to provide some intuition before we next discuss bulk tendencies.

**Bulk Tendencies**

*Shallow and Congestus Clouds*

3 km Separation

The first 4 panels of Fig. 10 show the bulk advective tendencies for parameters characteristic of shallow and mid-depth $KRoNUT_e$ within $L_s$. $L_s$ is 5 km, $H_s$ is 10 km, $w^*_s$ is 12 m s$^{-1}$, $w^*_e$ is 2 m s$^{-1}$, and $\Delta s$ is 3 km. The timestep range is 900 s-1200 s. Shallow circulations evolving under the influence of deep circulations always move toward the parent deep convection (Fig. 10a4) and always invigorate due to strong upward advection between the cores of the two circulations (Fig. 10a3). However, their geometric evolution depends strongly on initial geometry. The shortest and narrowest circulations grow vertically and horizontally, but most $KRoNUT_e$ deepen but narrow.

The third 4 panels of Fig. 10 show bulk diffusive tendencies; $v$ is 100 m$^2$ s$^{-1}$. The sign of the tendencies within any panel are less likely to be consistent than for advection, and overall, we observe sensitivity to numerics for the precise values calculated. Diffusion always acts to widen and deepen the shortest circulations likely due to low-level inflow to $KRoNUT_s$ at all levels of the evolving circulations's core of vorticity. However, taller clouds weakly narrow and shoal while a few are invigorated. (Fig. 10c). Except for $L_e \approx 3$ km, diffusion does not move the circulation center (Fig. 10c4 & 10d4).

8 km Separation

For the well-separated pair of KRoNUT, we will focus on differences with the adjacent pair (second and fourth 4 panels in Fig. 10). Broadly and unsurprisingly, advective tendencies are weaker when KRoNUT are 8 km apart than when they are 3 km apart. However, diffusive tendencies are of a similar magnitude. The biggest difference is that $\dot{H}_e$ is mostly negative due to advection. This is likely because KRoNUT$_e$ exists mostly within the subsident zone of KRoNUT$_s$ when $\Delta s$=8 km and $w_e$ is weakened in the core of the circulation in this case (Fig. 9d).

*Deep Clouds*

3 km Separation

For near-field deep circulations, $\Delta s$ is 3 km. Within the rising region of KRoNUT$_s$, the evolving flows grow upward and mostly contract inward (First 4 panels of Fig. 11). Most evolving flows shift toward the stationary flow, and for the shortest and widest flows, they do so with a velocity approaching the mesoscale at over 0.5 m s$^{-1}$. KRoNUT$_e$ with an $\alpha \approx 2$ are invigorated by KRoNUT$_e$.

Diffusive tendencies are rather weak (bottom left four panels of Fig. 11). Most circulations weaken consistent with Fig. 9e. The tallest and narrowest grow upward.

8km Separation

While advection tends to deepen and enervate a deep cloud nearby another, it tends to shoal (Fig. 11b1) and invigorate (Fig. 11b3) one far away. Interestingly, KRoNUT$_e$ are advected toward a common $L_e$ with the narrowest circulations widening and the widest narrowing. While tendency magnitudes are generally lower when clouds are farther apart, the separation tendency (Fig. 11b4) is of the same order as the nearby case. Diffusive tendencies of $H_e$ are nearly identical to those in the adjacent case (Fig. 11d1) while circulations are now weakly invigorated (consistent with Fig. 9f).

**Discussion and Conclusions**

In our development above, we have been careful not to conflate KRoNUT with clouds. In this section, we are less strict but will explicitly acknowledge that the DoNUT framework is yet incomplete.

When considering the bulk circulation tendencies collectively, it is worth noting the magnitude of $\dot{H}$ is an order of magnitude higher than the other parameter speeds. It is comparable to the wind speed used for the calculations (i.e. $w^*$). So, this suggests the notion that a cloud (and we might imagine a cloud top in the real world) rises at a fraction of its characteristic speed and that this fraction depends on the height of the cloud. Our conclusion is slightly different than Turner (1962) who suggests the former but that the latter should depend not on height but inversely on the aspect ratio. We show that isolated KRoNUT circulations tend to grow upward due to advection and diffusion (i.e. even in the absence of upward buoyancy).

If we accept a diffusive constant of $v$=100 m$^2$ s$^{-1}$ for the sake of argument, the remaining parameter tendencies separate into several behaviors. Small $L$ and small $H$ clouds (i.e. shallow convective clouds) are invigorated by advection, but this is largely counteracted by enervation from diffusion. They contract due to advection but widen due to diffusion almost equally. So, $L$ and $w^*$ may be steady while they grow upward. Small $L$ and large $H$ clouds (i.e. deep convective clouds) widen due to advection and diffusion, are weakened by diffusion, and grow strongly upward from advection. So, deep convective clouds evolve toward taller, wider, and weaker clouds from dry dynamics. This is how we frequently describe the maturation process of real-world clouds especially over tropical oceans. In our framework, tall and wide clouds

(perhaps, deep stratiform clouds) will still grow which is clearly unrealistic. But because of their geometric growth, they weaken with time toward a zero-energy state.

What about high *L* and low *H* clouds? Circulations in the bottom right corner of panels in Fig. 6, behave broadly similarly to other shallow clouds but with much stronger horizontal contraction tendency, although they do yield very different diffusive structure than other clouds (bottom row Fig. 7). It is unclear to the authors that this region of the phase space corresponds to any commonly observed cloud types. This could be because wide circulations rapidly contract and become narrow or it could be because they never form in the first place. Or, this region of the phase space may correspond to shallow stratiform clouds or many closely spaced, $\alpha \approx 1$, small *H* clouds. FIB25 also identify this region of the KRoNUT phase space as exhibiting unique behavior.

Perhaps the most surprising aspect of the details of KRoNUT dynamics is the range of behaviors possible between interacting circulations due solely to advection and diffusion. Adjacent clouds may attract or repel each other; they may invigorate or enervate the other; they may grow or shrink the other. The exercise in examining interacting clouds reenforces a broader conclusion that if one considers a cloud as an energetic entity then there are internal limits on how changes to the height, width, and intensity flow of the cloud, which each influence the total energy, may occur. Tendencies due to advection tend to depend most strongly on circulation height while tendencies due to diffusion are more likely to depends most strongly on circulation width. While FIB25 frame their arguments somewhat more precisely in terms of circulation density instead of energy, their conclusions in this regard are the same.

Though the KRoNUT seems a useful tool, it is clearly an incomplete description of convective flows. Comparisons to RAMS simulations remind us that downdrafts alter the nature of strong curvature of the flow near the surface in mature storms and that the tropopause limits the depth of outflow and vertical growth. These affect the nature of the dynamics of the RAMS circulations and yield mismatches with some KRoNUT dynamics. As we develop the model, we are actively working to minimize these mismatches and to add physics so that the DoNUT may better reflect the behavior of observed, diabatic clouds. The utility of the KRoNUT is its simple-but-not-too-simple conception which allows for analytic development (FIB25) and detailed interpretation (above).

**Code and Data Availability**

RAMS is available at https://github.com/RAMSmodel/RAMS. Simulation data are available upon request.

**Acknowledgements**

This work was supported by NSF ASG through Award 2224293 and by the UC Davis Academic Senate. MI is grateful to M. Woody for his helpful contribution.

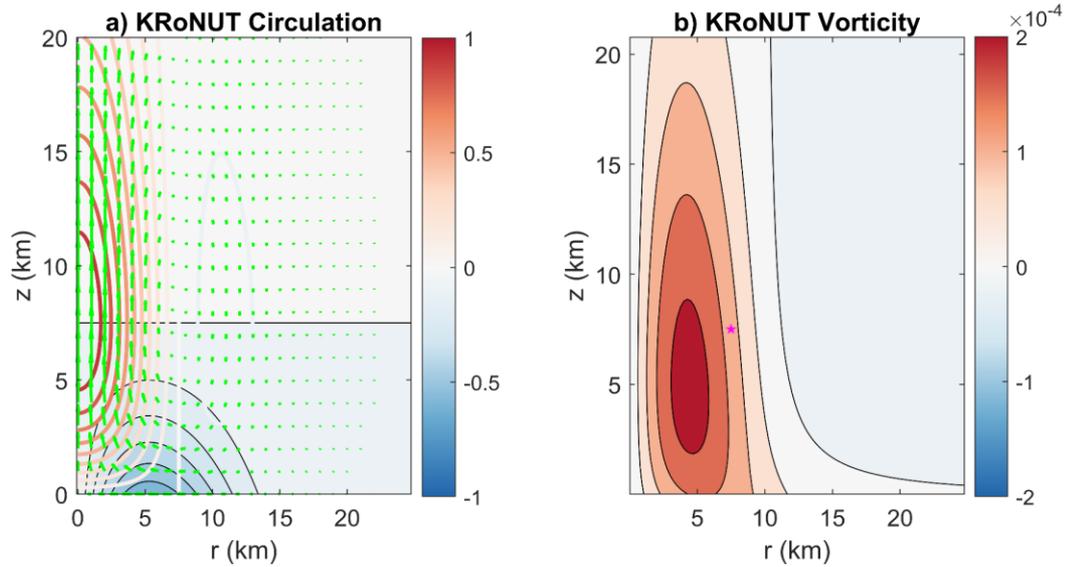

Figure 1: Radial cross-sections of KRoNUT kinematics. a) The KRoNUT velocity field. Horizontal wind is shown in filled black contours. The vertical wind is shown in open color contours. The total winds is shown in green vectors. Units are m s$^{-1}$. b) The KRoNUT vorticity field. Units are s$^{-1}$. Cross-sections are shown for $L$=7.5 km and $H$=7.5 km (marked with a magenta star).

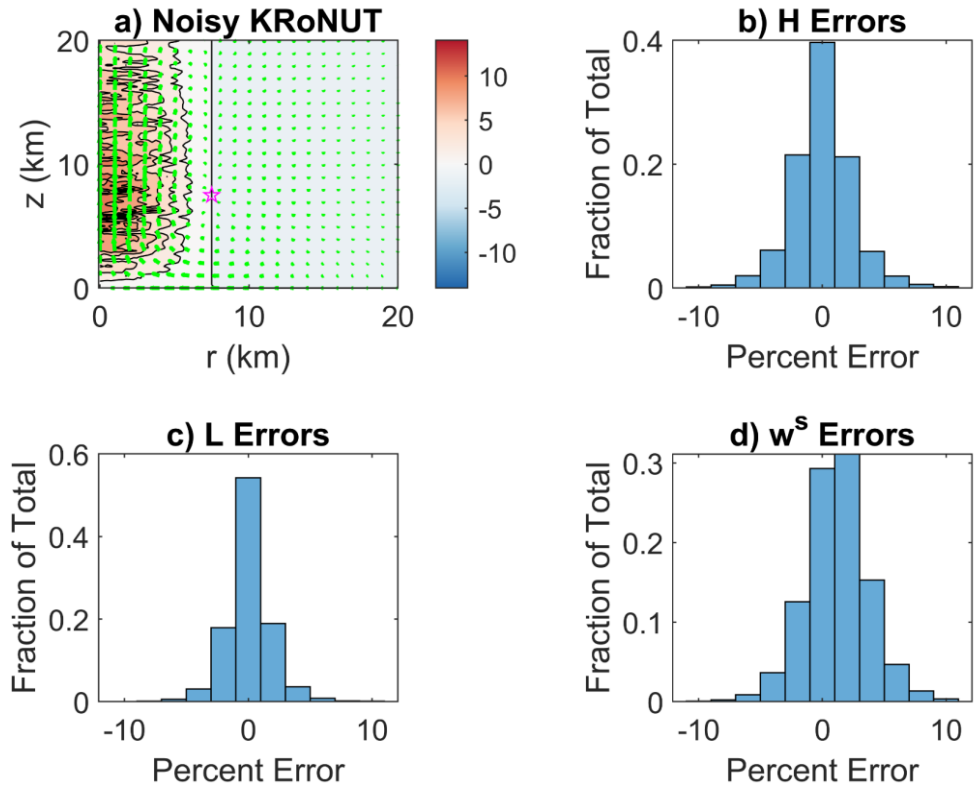

Figure 2: Error in fitting noisy KRoNUTs. a) An example of the vertical velocity field after the application of noise in filled contours (filled contours) and the total wind in the green vectors. b) – d) show the distributions of parameters errors among the 10000 iterations.

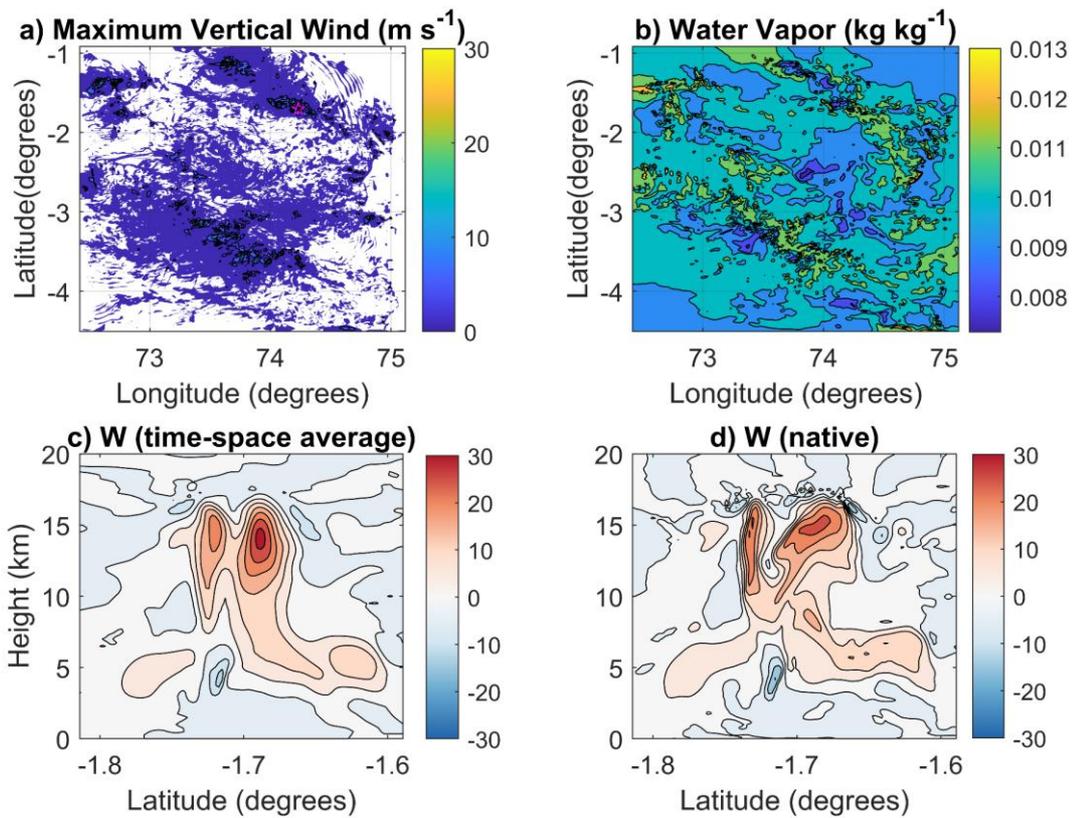

Figure 3: Output from the RAMS simulation at 4Z. a) The maximum vertical updraft within the free troposphere with negative values screened (to white). The magenta star marks the locations of the cross sections. b) The water vapor mixing ratio at 3km. c) & d) Cross sections of vertical wind speed when averaged over the preceding 5 minutes and to a standard 250m grid in c) and at the 4Z timestep on the native vertically stretched grid in d).

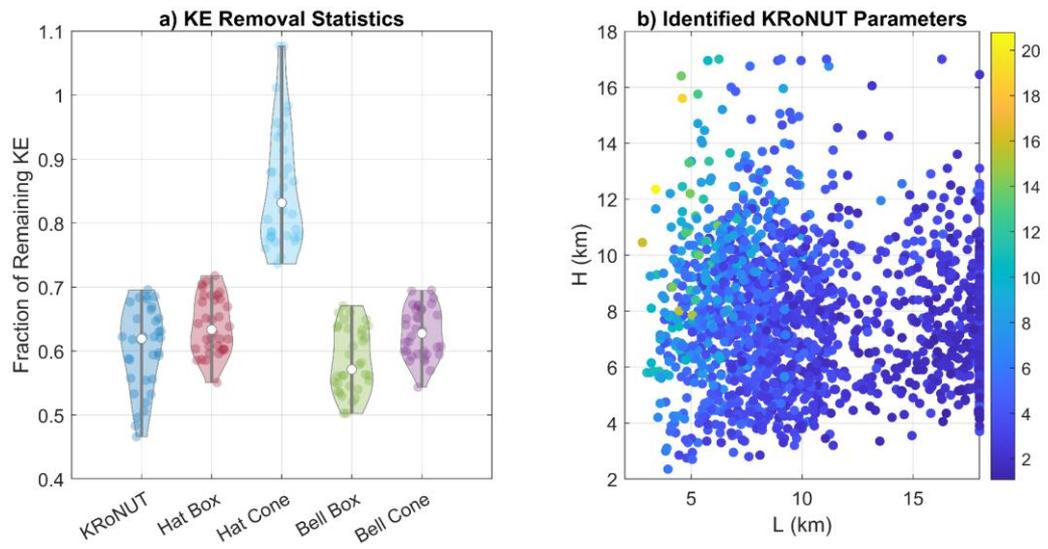

Figure 4: Results from fitting circulations in the RAMS simulations. a) The mean fraction of vertical Kinect energy remaining after 50 circulations have been removed across all 35 output times. b) Scatter of fit KRoNUT parameters. Color shading indicates the diagnosed $w^*$.

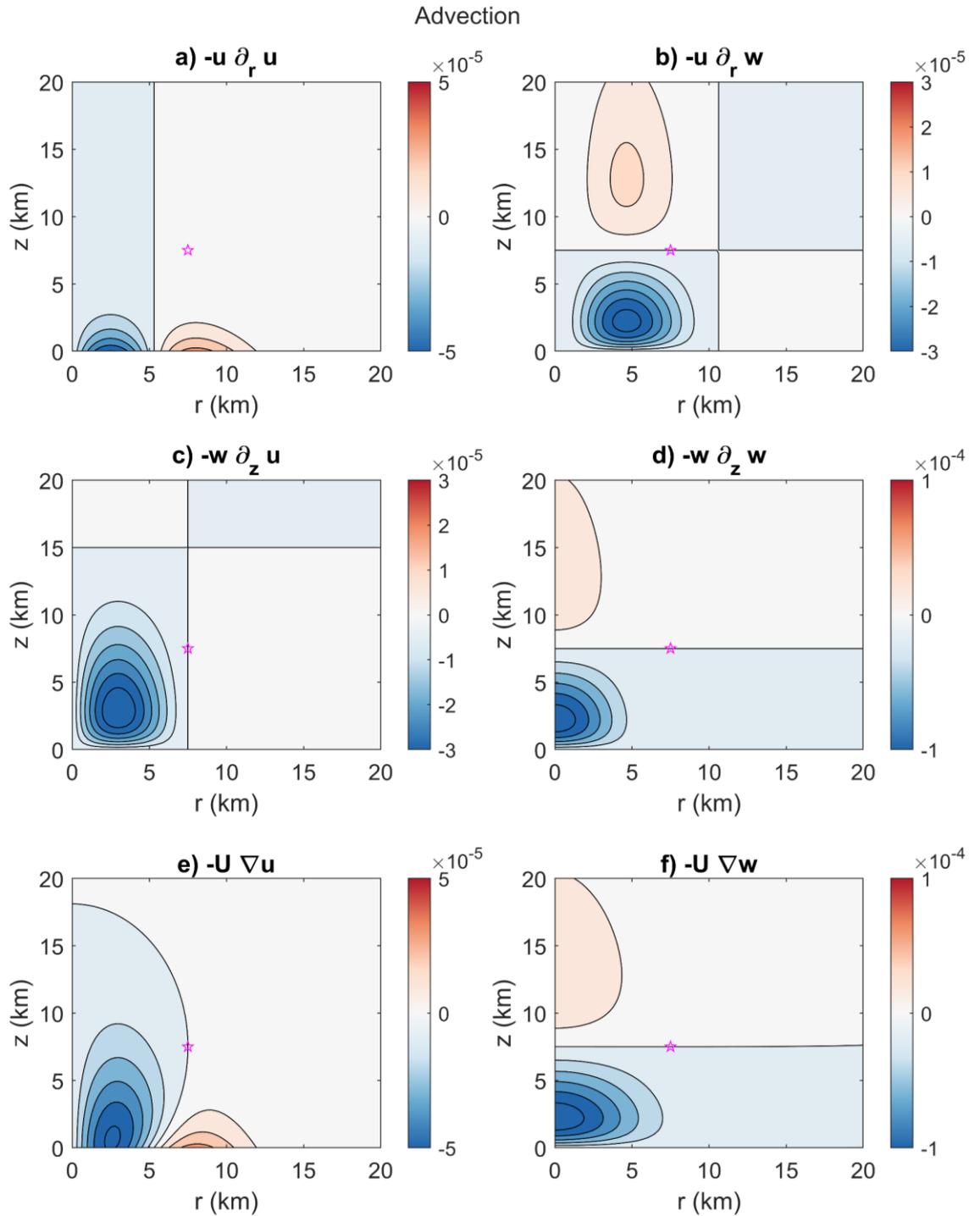

Figure 5: Radial cross-sections of advective terms. The left column shows the two components of the radial velocity forcing and the sum in e). The right column shows the two components of the vertical velocity and the sum in f). Magenta stars highlight the KRoNUT stagnation point at $L$, and $H$. Units are m s$^{-2}$.

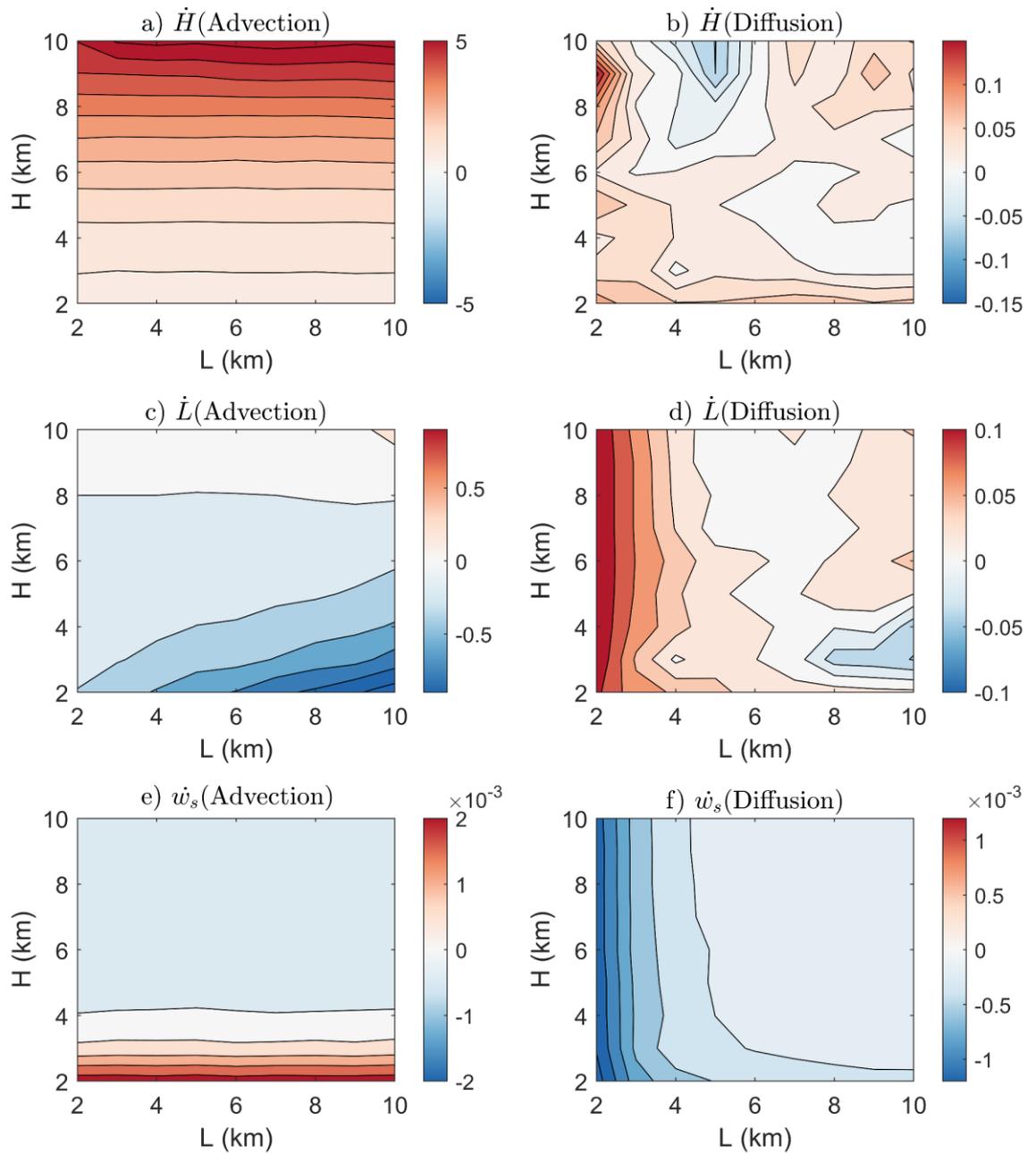

Figure 6: Parameter tendencies from advection (left column) and diffusion (right column). The units for the first two rows are m s$^{-1}$. The units for the last row are m s$^{-2}$.

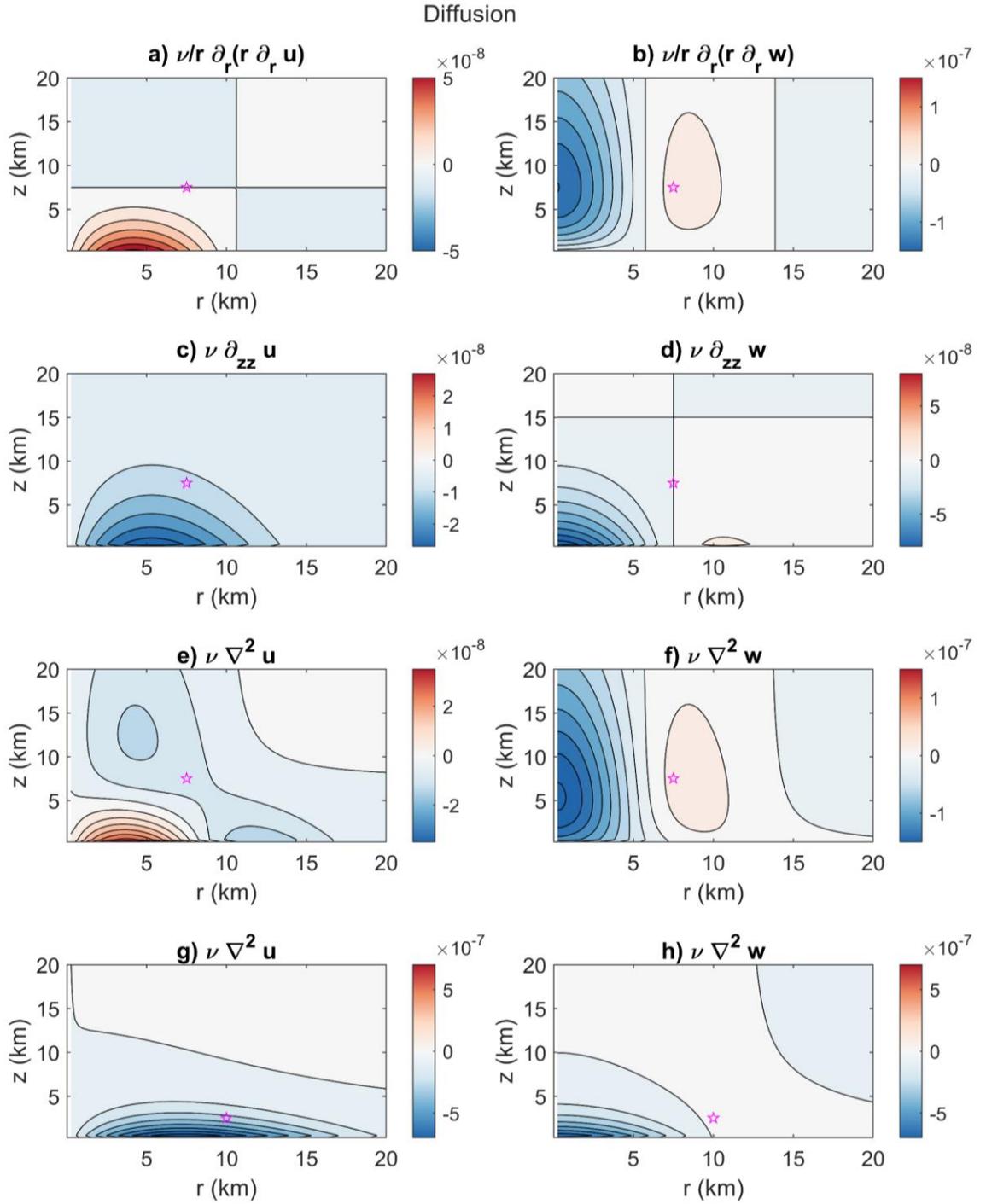

Figure 7: The figure is organized as Fig. 6 but for diffusive tendencies (to within a factor of $\nu$). Units of the cross-sections are m$^{-1}$ s$^{-1}$ (i.e. m s$^{-2}$ m$^2$ s$^1$). Note that the last row has different $L$ and $H$.

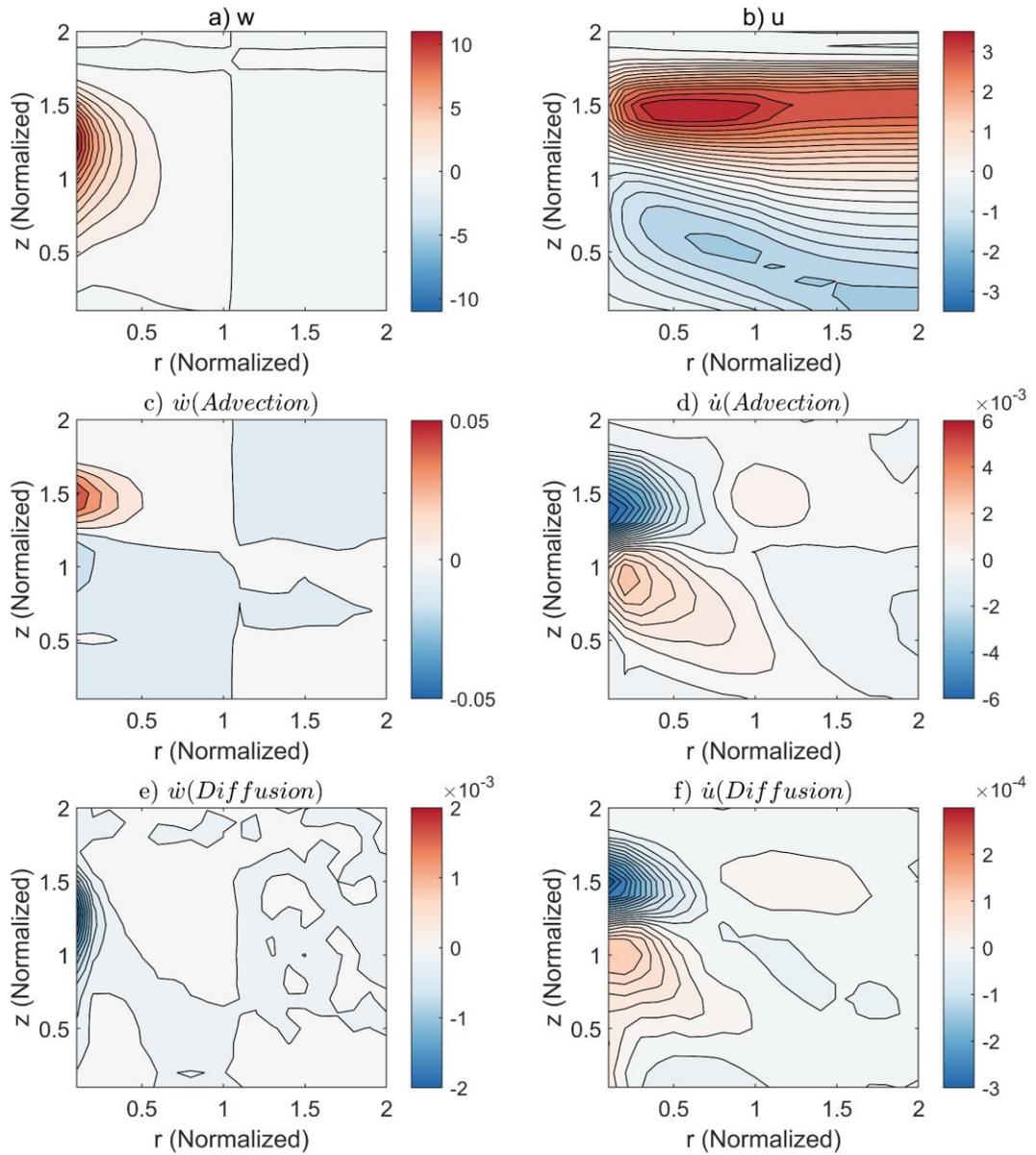

Figure 8: Composites from RAMS. Distances are normalized to fit *L* and *H* values. The first row shows kinematic composites, and the final two rows show dynamic composites from the dynamical tendencies computed in RAMS.

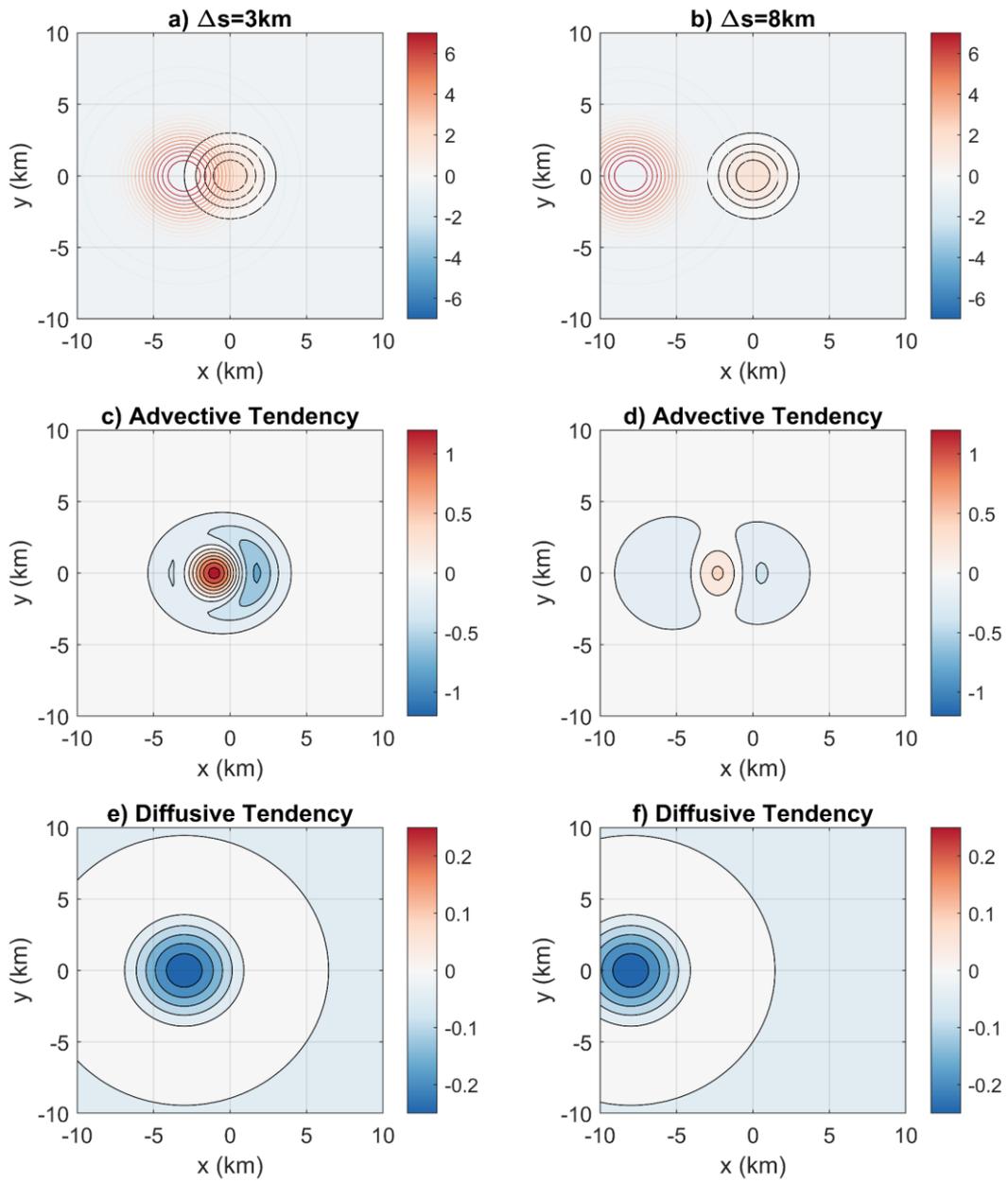

Figure 9: Vertical mean vertical velocities and vertical velocity tendencies for a pair of KRoNUT separated by 3 km (left column) and 8 km (right column). The first row shows the mean vertical velocity of the evolving KRoNUT in filled contours and the static KRoNUT in color contours. The final four panels show mean tendency of the vertical wind from advection (middle row) and diffusion (bottom row).

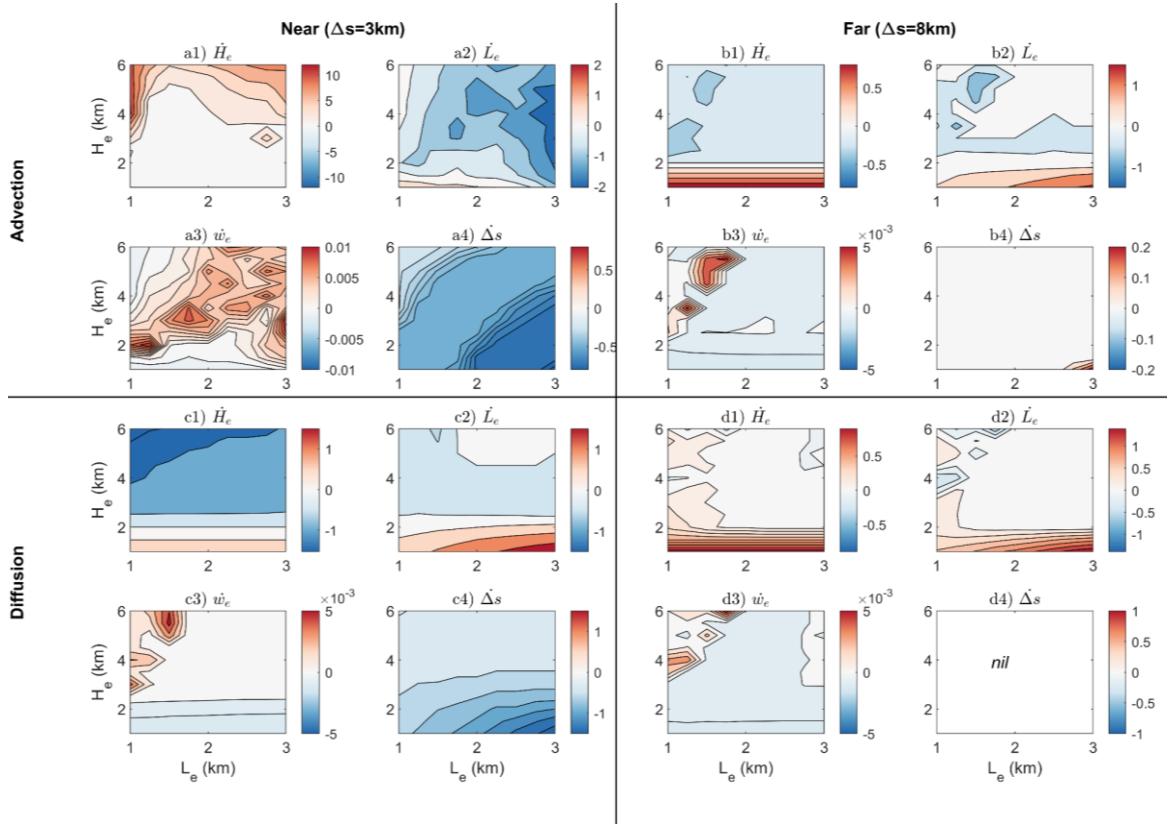

Figure 10: Parameter tendencies for the interacting deep (KRoNUT$_s$) and shallow (KRoNUT$_e$) circulation. The figure is divided into four quadrants. The left quadrants (a and c) show tendencies when the KRoNUT are separated by 3km. The rights quadrants (b and d) show tendencies when the KRoNUT are separated by 8km. The upper quadrants (a and b) show tendencies due to advection. The lower quadrants (c and d) show tendencies due to diffusion. Units of velocities (subpanels 1, 2, and 4) are m s$^{-1}$. Units of acceleration (subpanel 3) are m s$^{-2}$.

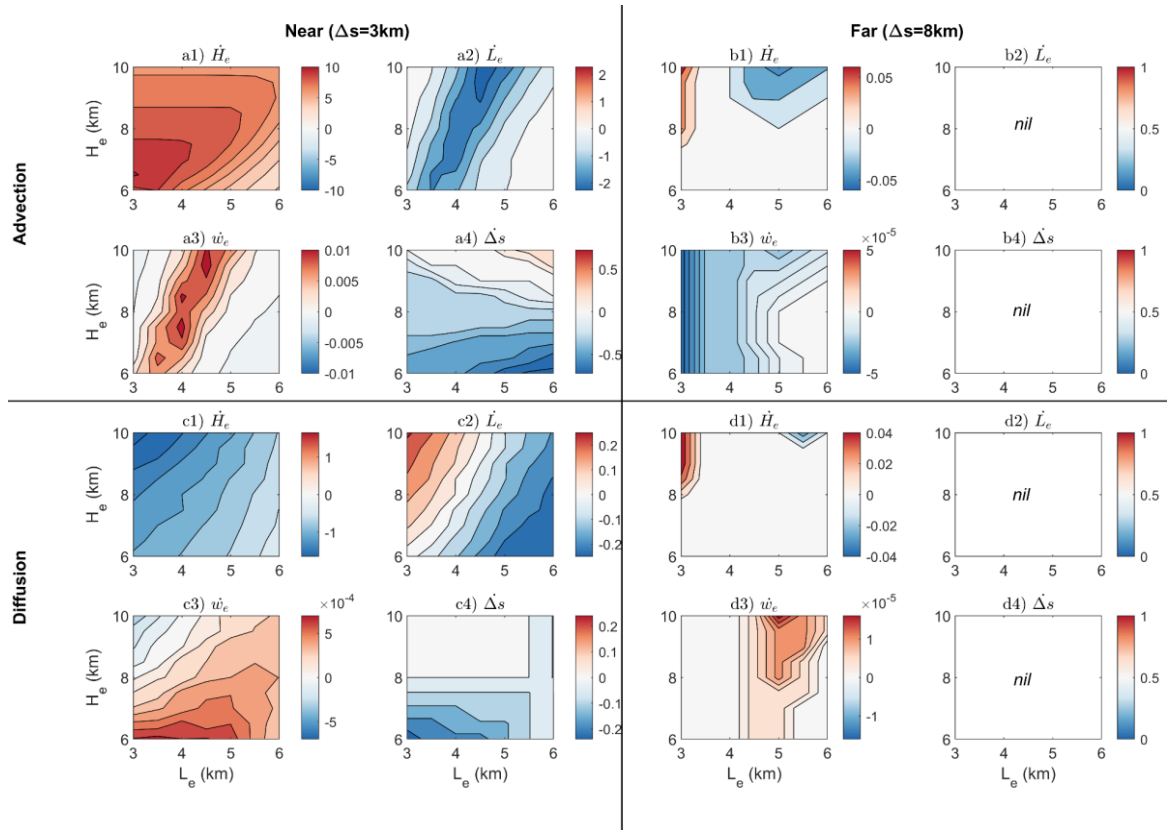

Figure 11: Same as Fig. 10 but for the case where KRoNUT$_e$ is a deep cloud. All values in b4) and d4) are 0 m s$^{-1}$.